\documentclass[superscriptaddress,twocolumn,showpacs,preprintnumbers]{revtex4}%
\usepackage{graphicx}
\usepackage{dcolumn}
\usepackage{bm}
\usepackage{latexsym}
\usepackage{amsmath}
\usepackage{amssymb}
\begin{document}
\title{Error correction during DNA replication} 
\author{Ajeet K. Sharma}
\author{Debashish Chowdhury{\footnote{Corresponding author(E-mail: debch@iitk.ac.in)}}}
\affiliation{Department of Physics, Indian Institute of Technology,
Kanpur 208016, India.}
\begin{abstract}
DNA polymerase (DNAP) is a dual-purpose enzyme that plays two opposite
roles in two different situations during DNA replication. It plays its
normal role as a {\it polymerase} catalyzing the elongation of a new DNA
molecule by adding a monomer. However, it can switch to the role of an
{\it exonuclease} and shorten the same DNA by cleavage of the last
incorporated monomer from the nascent DNA. Just as misincorporated
nucleotides can escape exonuclease causing replication error, correct
nucleotide may get sacrificed unnecessarily by erroneous cleavage.
The interplay of polymerase and exonuclease activities of a DNAP is
explored here by developing a minimal stochastic kinetic model of DNA
replication. Exact analytical expressions are derived for a few key
statistical distributions; these characterize the temporal patterns
in the mechanical stepping and the chemical (cleavage) reaction.
The Michaelis-Menten-like analytical expression derived for the average
rates of these two processes not only demonstrate the effects of their
coupling, but are also utilized to measure the extent of {\it replication
error} and {\it erroneous cleavage}.
\end{abstract}

\maketitle
\section{Introduction}

DNA polymerase (DNAP) replicates a DNA molecule; the sequence of the
nucleotides, the monomeric subunit of DNA, on the product of
polymerization is dictated by that on the corresponding template DNA
through the Watson-Crick complimentary base-paring rule \cite{kornberg}.
DNAP moves step-by-step along the template strand utilizing chemical
energy input and, therefore, these are also regarded as a molecular
motor \cite{howard,kolofisher}.

An unique feature of DNAP is that it is a dual-purpose enzyme that
plays two opposite roles in two different circumstances during DNA
replication. It plays its normal role as a polymerase catalyzing the
{\it elongation} of a new DNA molecule. However, upon committing an
error by the misincorporation of a wrong nucleotide, it switches
its role to that of a exonuclease catalyzing the {\it shortening}
of the nascent DNA by cleavage of the misincorporated nucleotide at
the growing tip of the elongating DNA \cite{krantz10}.
The two distinct sites on the DNAP where, respectively, polymerization
and cleavage are catalyzed, are separated by 3-4 nm
\cite{ibarra09}. The nascent DNA
is transferred back to the site of polymerization after cleaving the
incorrect nucleotide from its growing tip. The elongation and cleavage
reactions are thus {\it coupled} by the transfer of the DNA between
the sites of polymerase and exonuclease activity of the DNAP. However,
the physical mechanism of this transfer is not well understood
\cite{xie09}.

In this paper we develop a minimal kinetic model of DNA replication
(more precisely, that of the ``leading strand'' which can proceed
continuously)
that captures the coupled polymerase and exonuclease activities of a
DNAP within the same theoretical framework. From this model, we
derive the exact analytical expressions for
(i) the dwell time distribution (DTD) of the DNAP at the successive
nucleotides on the template DNA, and (ii) the distribution of the
turnover times (DTT) of the exonuclease (i.e., the time intervals
between the successive events of cleavage of misincorporated
nucleotide from the nascent DNA). The mean of these two distributions
characterize the average rates of elongation and cleavage, repectively;
we show that both can be written as Michaelis-Menten-like expressions
for enzymatic reactions which reveals the effect of coupling explicitly.

In our model, the kinetic pathways available to the correct and incorrect
nucleotides are the same. However, it is the ratio of the rate constants
that makes a pathway more favorable to one species than to the other.
Similar assumption was made by Galas and Branscomb \cite{galas78} in one
of the earliest models of replication.
Therefore, in spite of the elaborate quality control system, some
misincorporated nucleotides can escape cleavage; such {\it replication
error} in the final product is usually about $1$ in $10^{9}$ nucleotides.
Moreover, occasionally a correct nucleotide is erroneously cleaved
unneccessarily; such ``futile'' cycles slow down replication
\cite{fersht82}.

We define quantitative measures of these two types of error and derive
their exact analytical expressions from our model for ``wild type'' DNAP.
Using special cases of these analytical expressions, we also examine
the effects of two different mutations of the DNAP \cite{ibarra09}-
(i) ``exo-deficient'' mutant that is incapable of exonuclease activity,
and (ii) ``transfer-deficient'' mutant on which the rate of transfer
to the exonuclease site is drastically reduced.

\section{Model}

\begin{figure}
\includegraphics[angle=0,width=0.9\columnwidth]{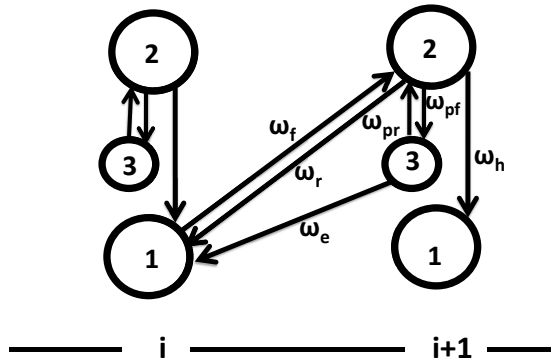}
\caption{The kinetic model of DNA replication. The chemo-mechanical
states of a single  DNAP and the allowed transitions are shown along
with the corresponding transition probabilities per unit time (the
rate constants). The corresponding rate constants for the incorrect
nucleotide are denoted by the symbol $\Omega$ (see the text for
details).}
\label{fig-3state}
\end{figure}

Almost all DNAP share a common ``right-hand-like'' structure.
Binding of the correct dNTP substrate triggers closing of the ``hand''
which is required for the formation of the diester bond between the
recruited nucleotide monomer and the elongating DNA molecule.
The kinetic scheme of our stochastic model of replication is shown in
fig (\ref{fig-3state}). The rate constants for the correct and incorrect
nucleotides are denoted by $\omega$ and $\Omega$, respectively; the
same subscript is used in both the cases for the same transition.

Let us begin with the situation where the DNAP is ready to begin its
next elongation cycle; these mechano-chemical state is labelled by the
integer index $1$. In principle, the transition $1 \to 2$ consists of
two steps: binding of the dNTP substrate and the formation of the
diester bond. The overall rate of this step is $\omega_f$ for a 
correct substrate and $\Omega_{f}$ for an incorrect substrate. 

Occasionally, because of the random fluctuation of the ``hand'' between
the ``open'' and ``closed'' conformations, the dNTP may escape even
before the formation of the diester bond; this takes place with the rate
constant $\omega_r$. If the recruited dNTP is incorrect, the hand remains
``open'' most of the time and the rate constant for the rejection of the
dNTP is $\Omega_r$ ($\Omega_r >> \omega_r$).
Note that dNTP selection through $1 \rightleftharpoons 2$ involves a
discrimination between the correct and incorrect dNTP substrate
on the basis of free energy gained by complementary base-pairing with
the template.

The transition $2(i+1) \to 1(i+1)$ corresponds to the {\it relaxation}
of the freshly incorporated nucleotide to a conformation that allows
the DNAP to be ready for the next cycle. The rate constants for this
step are $\omega_{h}$ and $\Omega_{h}$ respectively, for correctly
and incorrectly incorporated nucleotides. Alternatively, while in the
state $2(i+1)$, the DNAP can transfer the growing DNA molecule to its
exonuclease site; this transfer takes place at a rate $\omega_{pf}$
($\Omega_{pf}$) if the selected nucleotide is correct (incorrect).
Since $\Omega_h << \omega_h$, and $\Omega_{pf} >> \omega_{pf}$, the
misincorporated nucleotide most often gets transferred to the
exonuclease site whereas relaxation, rather than transfer, is the
most probable pathway when the incorporated nucleotide is correct.

The actual cleavage of the diester bond that severs the nucleotide at
the growing tip of the DNA is represented by the transition
$3(i+1) \to 1(i)$. For a correct nucleotide, $\omega_{pr} >> \omega_{e}$
indicating that the DNA is likely to be transferred back to the polymerase
site without the unneccesary cleavage of the correct nucleotide.
In contrast, for an incorrect nucleotide, $\Omega_{pr} << \Omega_{e}$
which makes error correction a highly probable event. Moreover,
$\Omega_{pr} << \omega_{pr}, \Omega_{e} >> \omega_{e}$.

Since the trimmed DNA is transferred to the polymerase site extremely
rapidly \cite{krantz10}, each of the rate constants $\omega_{e}$ and
$\Omega_{e}$ incorporate both the trimming and transfer.

\begin{figure}[td]
\begin{center}
\includegraphics[width=0.4\columnwidth]{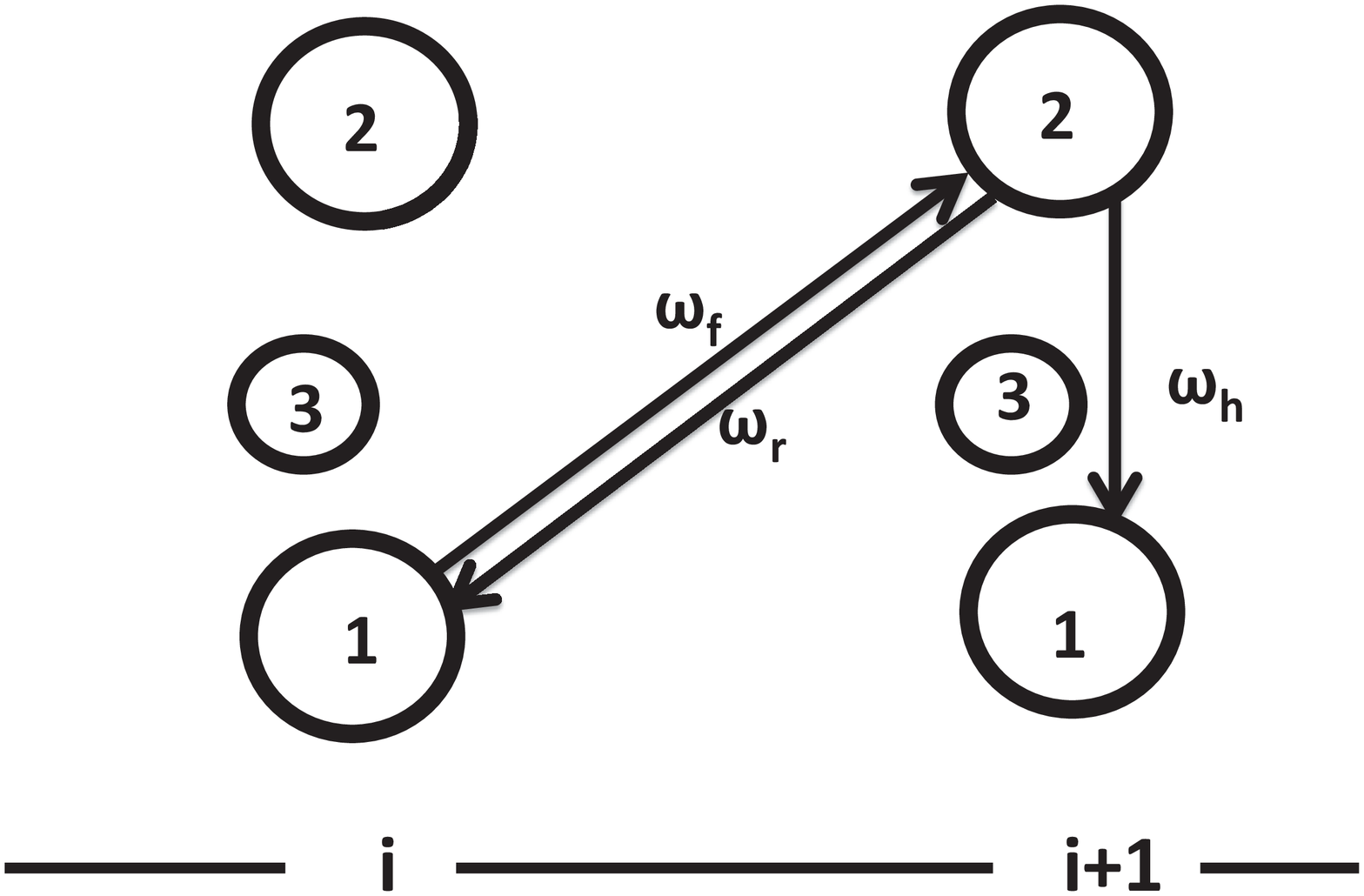}
\includegraphics[width=0.4\columnwidth]{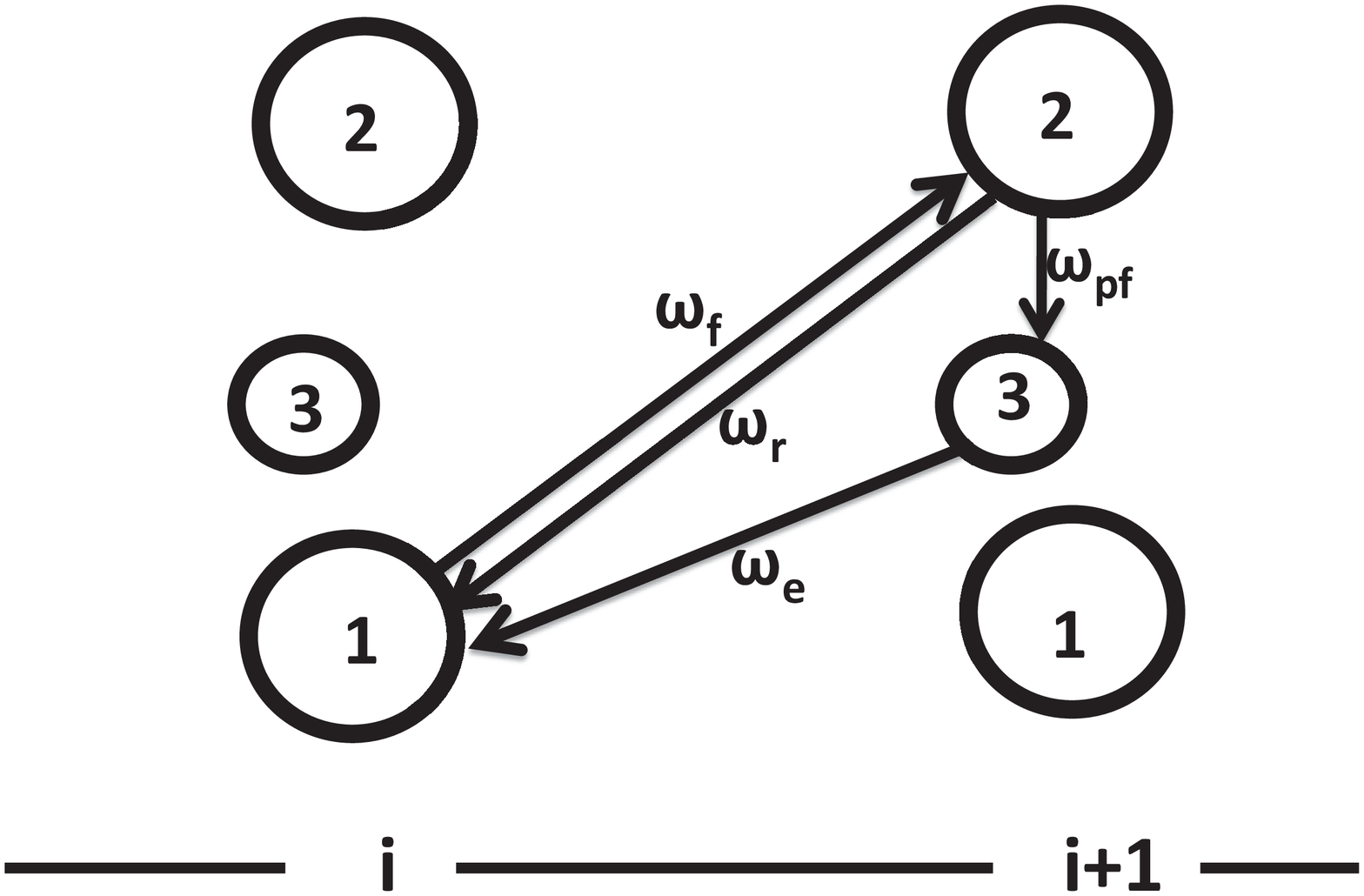}
\end{center}
\caption{``Purely'' polymerizing reaction (left panel) and
``pure'' exonuclease reaction (right panel).}
\label{fig-MMlikeP}
\end{figure}

Interestingly, the full kinetic scheme in fig.\ref{fig-3state} can be
viewed as a {\it coupling} of a purely polymerase-catalyzed reaction
(shown in the left panel of fig.\ref{fig-MMlikeP})
and a purely exonuclease-catalyzed reaction
(shown in the right panel of fig.\ref{fig-MMlikeP});
the transition with the rate $\omega_{pr}$ couples these two reactions.

Strictly speaking, for an ``exo-deficient'' DNAP
\cite{ibarra09},
$\omega_{e} = 0 = \Omega_{e}$ although the rates of forward
and reverse transfer between the sites of polymerase and
exonuclease activities may not be necessarily negligible.
Similarly, either  $\omega_{pf} = 0 = \Omega_{pf}$,
or $\omega_{pr} = 0 = \Omega_{pr}$ (or, both) can be the
cause of ``transfer-deficiency'' of the DNAP.

\section{Results and discussion}

\subsection{Distribution of Dwell time}

The DTD considered here arises from intrinsic stochasticity and not
caused by any sequence inhomogeneity of the mRNA template \cite{schwartz11}.
For a molecular motor that is allowed to step backward as well as forward, 
we use positive ($+$) and negative ($-$) signs to represent the forward 
and backward steps, respectively. For example, $\psi_{+-}(t)$ is the 
{\it conditional} DTD (cDTD) \cite{garai11} when a forward step is 
followed by a backward step and $p_{+-}$ is the probability of such a 
transition. Therefore, the DTD can be written as
\begin{equation}
\psi(t)=p_{+-}\psi_{+-}(t)+p_{++}\psi_{++}(t)+p_{-+}\psi_{-+}(t)+p_{--}\psi_{--}(t)
\end{equation}
Since in our model two consecutive backward steps are forbidden implies that 
$p_{--}\psi_{--}(t)=0$. 
We calculate the cDTD following the standard method \cite{chemla} 
that has been used successfully earlier for the calculation of 
cDTD for some other motors (see, for example, ref.\cite{garai11}).

Let $P_{\mu}(j,t)$ be the probability of finding the DNAP in the $\mu$-th 
($\mu=1,2,3$) chemical state at the $j$-th site (i.e., at the discrete 
position $x_j$ ($j = \infty,...-1,0,1,...\infty$). Then master equations 
for $P_{\mu}(j,t)$ are  
\begin{eqnarray}
\dfrac{dP_1 (j,t)}{dt}&=&-\omega_f P_1 (j,t) + \omega_r P_2 (j+1,t)+\omega_h P_2 (j,t)\nonumber \\
&+& \omega_e P_3 (j+1,t)
\label{eq-meq1}
\end{eqnarray}
\begin{eqnarray}
\dfrac{dP_2 (j,t)}{dt}&=&\omega_f P_1 (j-1,t) - (\omega_r+\omega_{pf}+\omega_h)P_2 (j,t)\nonumber\\
&+&\omega_{pr} P_3 (j,t)
\label{eq-meq2}
\end{eqnarray}
\begin{equation}
\dfrac{dP_3 (j,t)}{dt}=\omega_{pf} P_2 (j,t) - (\omega_e+\omega_{pr})P_3 (j,t)
\label{eq-meq3}
\end{equation}
In terms of the Fourier transform
\begin{equation}
\bar{P}_{\mu}(q,t) = \sum_{j=-\infty}^{\infty} P_{\mu}(x_{j},t) e^{-iqx_j}
\label{eq-FourierP}
\end{equation}
of $P_{\mu}(x_{j},t)$, the master equations can be written as a
matrix equation
\begin{equation}
 \dfrac{d}{dt}\bar{\textbf{P}}(q,t)=\textbf{M(q)}\bar{\textbf{P}}(q,t)
 \label{meq}
\end{equation}
where $\bar{\textbf{P}}(q,t)$ is a column vector whose 3 components are  
$\bar{P}_{1}(q,t),\bar{P}_{2}(q,t),\bar{P}_{3}(q,t)$ and 
\begin{equation}
\textbf{M(q)}=
\begin{bmatrix}
  -\omega_f & \omega_h+\omega_r \rho_{-}(q) & \omega_e \rho_{-}(q) \\
   \omega_f\rho_{+}(q) & -(\omega_h+\omega_r+\omega_{pf}) & \omega_{pr} \\
   0 & \omega_{pf} &-(\omega_e+\omega_{pr}) 
\end{bmatrix} 
\label{eq-Mq}
\end{equation}
with $\rho_{+}(q)=e^{-iqd}$ and  $\rho_{-}(q)=e^{iqd}$; $d$ being 
the step size, ie., $x_{j+1}-x_{j}=d$.

Taking Laplace transform of (\ref{meq}) with respect to time
\begin{equation}
\tilde{P}_{\mu}(q,s) = \int_{0}^{\infty} \bar{P}_{\mu}(q,t) e^{-s t},
\end{equation}
the solution of the master equation in the Fourier-Laplace space is 
\begin{equation}
\tilde{{\bf P}}(q,s) = {\bf R}(q,s)^{-1} \tilde{{\bf P}}(0)
\end{equation}
where 
\begin{equation}
{\bf R}(q,s) = s{\bf I} - {\bf M}(q) 
\end{equation}
and $\tilde{{\bf P}}(0)$ is the column vector corresponding to the 
initial probabilities.

Now we define
\begin{equation}
\tilde{P}(q,s)=\sum\limits_{i=1}^3 \tilde{P}_{i}(q,s)
\end{equation}
which can be calculated from 
\begin{equation}
\tilde{P}(q,s)=\dfrac{\sum\limits_{i,j=1}^3 Co_{j,i}P_{j}(0)}{\lvert{\textbf{R}}(q,s)\rvert}
\label{tprob}
\end{equation}
where $Co_{j,i}$ are the cofactors of the {\textbf{R}}(q,s).

The determinant of the matrix {\textbf{R}}(q,s) is a third order polynomial 
of $s$ and can be expressed as 
\begin{equation}
\lvert{\textbf{R}}(q,s)\rvert=s^3+\alpha s^2+\beta(q)s+\gamma (q)
\end{equation}
Note that $\alpha$ is independent of $q$ whereas $\beta$ and $\gamma$ 
are the function of $q$. For the explicit form (\ref{eq-Mq}) of ${\bf M}$ 
the coefficients $\alpha$, $\beta(q)$ and $\gamma(q)$ are given below.
\begin{equation}
\alpha=\omega_e+\omega_f+\omega_h+\omega_{pf}+\omega_{pr}+\omega_{r}
\end{equation}
$\beta$ can be expressed as
\begin{equation}
\beta(q)=\beta(0)+\beta_{+}(1-\rho_{+}(q))+\beta_{+-}(1-\rho_{+}(q)\rho_{-}(q))
\end{equation}
where
\begin{eqnarray}
\beta(0)&=&\omega_{e}\omega_{f}+\omega_{e}\omega_{h}+\omega_{e}\omega_{pf}+\omega_{f}\omega_{pf}+\omega_f\omega_{pr}\nonumber\\
&+&\omega_{h}\omega_{pr}+\omega_{e}\omega{r}+\omega_{pr}\omega_{r} 
\label{eq-beta0}
\end{eqnarray}
\begin{equation}
\beta_{+}=\omega_{f}\omega_{h}
\end{equation}
and
\begin{equation}
\beta_{+-}=\omega_{f}\omega_{r}. 
\end{equation}
Similarly, 
\begin{equation}
\gamma(q)=\gamma_{+}(1-\rho_{+}(q))+\gamma_{+-}(1-\rho_{+}(q)\rho_{-}(q)) 
\label{eq-gammaq}
\end{equation}
where
\begin{equation}
\gamma_+=\omega_{e}\omega_{f}\omega_{h}+\omega_{f}\omega_{h}\omega_{pr}
\end{equation}
and
\begin{equation}
\gamma_{+-}=\omega_{e}\omega_{f}\omega_{pf}+\omega_{e}\omega_{f}\omega_{r}+\omega_{f}\omega_{pr}\omega_{r}
\end{equation}

For convenience, we define the $2 \times 2$ diagonal matrix 
\begin{equation}
\mathbf{\rho(q)}=
\begin{bmatrix}
    \rho_{+}(q) & 0 \\
     0 & \rho_{-}(q)
\end{bmatrix}
\end{equation}
the column vector 
\begin{equation}
\mathbf{\Psi(s)}=\dfrac{1}{s}
\begin{bmatrix}
    1-p_{++}\psi_{++}(s)-p_{+-}\psi_{+-}(s) \\
    1-p_{-+}\psi_{-+}(s)-p_{--}\psi_{--}(s)\\
\end{bmatrix}
\end{equation}
and the $2 \times 2$ matrix 
\begin{equation}
\mathbf{\psi(s)}=
\begin{bmatrix}
      p_{++}\psi_{++}(s) & p_{+-}\psi_{+-}(s) \\
      p_{-+}\psi_{-+}(s) & p_{--}\psi_{--}(s)\\
\end{bmatrix}
\end{equation}
where $\psi_{\pm\pm}(s)$ are the Laplace transforms of the cDTDs 
$\psi_{\pm\pm}(t)$. 

$\tilde{P}(q,s)$ and cDTD are related \cite{chemla} by the equation
\begin{equation}
\tilde{P}(q,s)=  \textbf{p}_{0}^{T}( \textbf{I}- \mathbf{\psi}(s) \mathbf{\rho}(q))^{-1} \mathbf{\Psi}(s)
\label{m20}
\end{equation}
where $\textbf{p}_{0}$ is the vector of initial conditions. For 
example, $\textbf{p}_{0}^{T} = (1~ 0)$ corresponds to the given 
condition that the motor has taken the initial step in the forward 
(+) direction.

Thus, in principle, if one can calculate $\tilde{P}(q,s)$, one can use 
the relation (\ref{m20}) to solve for $\psi_{\pm\pm}(s)$ and, then taking 
inverse Laplace transform, obtained $\psi_{\pm\pm}(t)$. 
To calculate $\tilde{P}(q,s)$, one has to use an appropriate set of 
initial conditions consistent with the defnition of the dwell times. 
The set $P_{1}=0$,$P_{2}=1$,$P_{3}=0$ ensures that first step is taken 
forward. In other words, in our calculation, we start  the clock by 
setting it to $t=0$ when the DNAP reaches the state $2$ at $j$ from 
state $1$ at $j-1$. Therefore, in this case $p_{--}=0=p_{-+}$.
Corresponding to this initial condition, we now define,  
\begin{equation}
\tilde{P}_{+}(q,s)=\left.\tilde{P}(q,s)\right|_{\{P_{1}(0)=0,P_{2}(0)=1,P_{3}(0)=0\}}
\end{equation}
and, from equation (\ref{m20}), we get \cite{chemla}
\begin{equation}
\left.\dfrac{1}{s\tilde{P}_{+}(q,s)}\right|_{\{\rho_{-}(q)=0\}}=\dfrac{1-\rho_{+}(q)p_{++}\tilde{\psi}_{++}(s)}{1-p_{++}\tilde{\psi}_{++}(s)-p_{+-}\tilde{\psi}_{+-}(s)}
\label{m25}
\end{equation}
Equation (\ref{m25}) can be re expressed as
\begin{equation}
\left.\dfrac{1}{s\tilde{P}_{+}(q,s)}\right|_{\{\rho_{-}(q)=0\}}=a_0+a_+\rho_+(q)
\label{m26}
\end{equation}
where 
\begin{eqnarray} 
a_0 &=& \dfrac{1}{1-p_{++}\tilde{\psi}_{++}(s)-p_{+-}\tilde{\psi}_{+-}(s)} \nonumber \\
a_+ &=& - \dfrac{p_{++}\tilde{\psi}_{++}(s)}{1-p_{++}\tilde{\psi}_{++}(s)-p_{+-}\tilde{\psi}_{+-}(s)}
\label{a0a+}
\end{eqnarray}
Hence,
\begin{equation}
p_{++}\tilde{\psi}_{++}(s)=-\frac{a_{+}}{a_{0}}
\label{m27}
\end{equation}
and 
\begin{equation}
p_{+-}\tilde{\psi}_{+-}(s)=\frac{a_{0}+a_{+}-1}{a_{0}}
\label{m27b}
\end{equation}
Therefore, next we obtain 
$\left.\dfrac{1}{s\tilde{P}_{+}(q,s)}\right|_{\{\rho_{-}(q)=0\}}$ 
directly from (\ref{tprob}) and, by comparing it with equation (\ref{m26}), 
find out the expressions for $a_{0}$ and $a_{+}$; substituting these 
expressions for $a_{0}$ and $a_{+}$ into equations (\ref{m27}) and 
(\ref{m27b}) we get 
$p_{++}\tilde{\psi}_{++}(s)$ and $p_{+-}\tilde{\psi}_{+-}(s)$, respectively.

Using the same initial condition, from equation (\ref{tprob}), we get
\begin{widetext}
\begin{eqnarray}
\tilde{P}_{+}(q,s)&=&\dfrac{s^2+s(\alpha-\omega_r(1-\rho_{-}(q))+\beta(0)-(1-\rho_{-}(q))(\omega_e\omega_{pf}+\omega_e\omega_r+\omega_{pr}\omega_r)}{s^3+\alpha s^2+\beta(q)s+\gamma(q)}\nonumber\\
\end{eqnarray}
Therefore,
\begin{equation}
\left.\dfrac{1}{s\tilde{P}_{+}(q,s)}\right|_{\{\rho_{-}(q)=0\}}  =\dfrac{s^3+\alpha s^2+s(\beta(0)+\beta_{+}+\beta_{+-})+\gamma_{+}+\gamma_{+-}-(s \beta_{+}+\gamma_{+})\rho_{+}(q)}{s^3+s^2(\alpha-\omega_r)+s(\beta(0)-(\omega_{e}\omega_{pf}+\omega_{e}\omega_{r}+\omega_{pr}\omega_{r}))}
\label{m29}
\end{equation}
Comparing equation (\ref{m29}) with equation (\ref{m26}) we identify $a_{0}$ 
and $a_{+}$ and substituting these expressions for $a_{0}$ and $a_{+}$ into 
(\ref{m27}) we get 
\begin{equation}
p_{++}\tilde{\psi}_{++}(s)=\dfrac{s\beta_{+}+\gamma_{+}}{s^3+\alpha s^2+s(\beta(0)+\beta_{+}+\beta_{+-})+\gamma_{+}+\gamma_{+-}}=\dfrac{s\beta_+ +\gamma_+}{(s+\omega_1)(s+\omega_2)(s+\omega_3)}
\label{m30}
\end{equation}
where $\omega_{1}$,$\omega_{2}$ and $\omega_{3}$ are roots of the following equation
\begin{equation}
\omega^3-\alpha \omega^2+\omega(\beta(0)+\beta_{+}+\beta_{+-})-(\gamma_{+}+\gamma_{+-})=0
\end{equation} 
Inverse Laplace transformation of equation (\ref{m30}) gives the exact expression of $p_{++}\psi_{++}(t)$
\begin{equation}
 p_{++}\psi_{++}(t)=\dfrac{e^{-\omega_{1}t}(\gamma_{+}-\beta_{+}\omega_{1})}{(\omega_{1}-\omega_{2})(\omega_{1}-\omega_{3})}+
\dfrac{e^{-\omega_{2}t}(\gamma_{+}-\beta_{+}\omega_{2})}{(\omega_{2}-\omega_{1})(\omega_{2}-\omega_{3})}+
\dfrac{e^{-\omega_{3}t}(\gamma_{+}-\beta_{+}\omega_{3})}{(\omega_{3}-\omega_{1})(\omega_{3}-\omega_{2})}
\end{equation}

Similarly, using the expressions of $a_{0}$ and $a_{+}$ in eq. (\ref{m27b}), 
we get 
\begin{equation}
p_{+-}\tilde{\psi}_{+-}(s)=\dfrac{s^2\omega_r+s(\beta_{+-}+\omega_{e}\omega_{pf}+\omega_{e}\omega_{r}+\omega_{pr}\omega_{r})+\gamma_{+-}}{s^3+\alpha s^2+s(\beta(0)+\beta_{+}+\beta_{+-})+\gamma_{+}+\gamma_{+-}} 
\label{m33}
\end{equation}
Inverse Laplace transformation gives the exact expression of $p_{+-}\psi_{+-}(t)$
\begin{equation}
 p_{+-}\psi_{+-}(t)=\dfrac{e^{-\omega_{1}t}(\gamma_{+-}-c_{1}\omega_{1}+\omega_{1}^2\omega_{r})}{(\omega_{1}-\omega_{2})(\omega_{1}-\omega_{3})}+
                    \dfrac{e^{-\omega_{2}t}(\gamma_{+-}-c_{1}\omega_{2}+\omega_{2}^2\omega_{r})}{(\omega_{2}-\omega_{1})(\omega_{2}-\omega_{3})}+
                    \dfrac{e^{-\omega_{3}t}(\gamma_{+-}-c_{1}\omega_{3}+\omega_{3}^2\omega_{r})}{(\omega_{3}-\omega_{1})(\omega_{3}-\omega_{2})}
\end{equation}

\end{widetext}

\begin{figure}[t]
\begin{center}
\includegraphics[angle=-90,width=0.9\columnwidth]{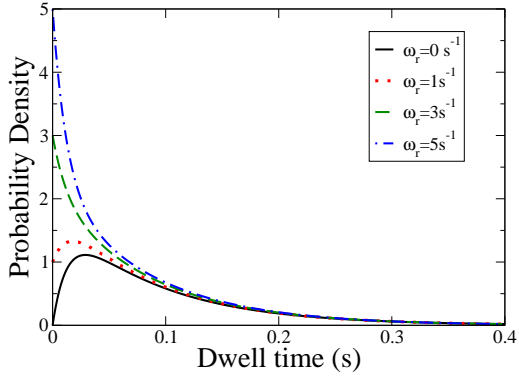}
\end{center}
\caption{(Color online) Probability density of conditional dwell time, 
$\psi_{+-}$(t) is plotted for a few different values of parameter 
$\omega_{r}$. The values of other parameters are (all in $s^{-1}$) 
$\omega_{f}$=20.0, $\omega_{pf}$=30.0, $\omega_{pr}$=15.0, 
$\omega_{h}$=40.0 and $\omega_{e}=$4.0. 
}
\label{+-}
\end{figure}

\begin{figure}
\begin{center}
\includegraphics[angle=-90,width=0.9\columnwidth]{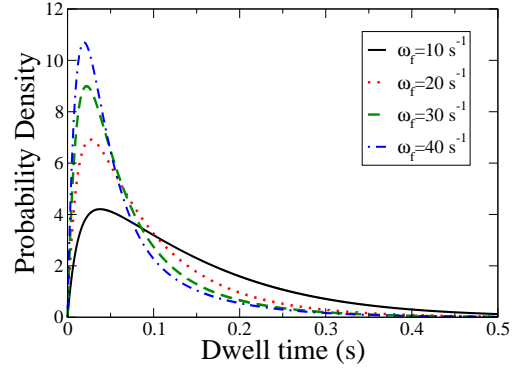}
\end{center}
\caption{(Color online) Probability density of conditional dwell time, 
$\psi_{++}$(t) is plotted for a few different values of parameter 
$\omega_{f}$. The values of other parameters are (all in $s^{-1}$) 
$\omega_{r}$=5.0, $\omega_{pf}$=30.0, $\omega_{pr}$=15.0, 
$\omega_{h}$=40.0 and $\omega_{e}=4.0$.}
\label{++}
\end{figure}

Note that by putting $s=0$ in equation (\ref{m30}) and (\ref{m33}) we get 
the ``branching probabilities''
\begin{equation}
p_{++}=\dfrac{\gamma_{+}}{\gamma_{+}+\gamma_{+-}}
\end{equation}
\begin{equation}
p_{+-}=\dfrac{\gamma_{+-}}{\gamma_{+}+\gamma_{+-}}
\end{equation}
which satisfy the normalization condition $p_{++}+p_{+-}=1$. 
The cDTDs $\psi_{+-}(t)$ and $\psi_{++}(t)$ are plotted in 
figs. \ref{+-} and \ref{++}, respectively, for a few different 
values of the parameters $\omega_{r}$ and $\omega_{f}$. 
In both the figures, the most probable dwell time increases 
with decreasing $\omega_{r}$ and decreasing $\omega_{f}$.  
 

In the same matrix-based formalism, the average velocity of a DNAP 
is given by the general expression \cite{chemla} 
\begin{equation}
V_{p}=- i \dfrac{\dot{\gamma}(0)}{\beta(0)},  
\label{eq-generalV} 
\end{equation} 
where the dot indicates derivative with respect to $q$.  
The right hand side of eqn.(\ref{eq-generalV}) can be evaluated for 
our model of DNAP using the explicit expressions (\ref{eq-gammaq}) 
and (\ref{eq-beta0}) for $\gamma(q)$ and $\beta(0)$, respectively. 
For the purpose of showing close relation of $V$ with the Michaelis-Menten 
(MM) equation for the average rates of enzymatic reaction, we now assume 
that dNTP binding is rate limiting (the general framework of our theory 
does not need this assumption). Under this assumption, we can write
\begin{equation}
\omega_{f}=\omega_{f}^0[dNTP_c]~{\rm and}~ \Omega_{f}=\Omega_{f}^0[dNTP_w]
\label{eq-concdep}
\end{equation}
where $[dNTP_c]$ and $[dNTP_w]$ are the concentrations of the correct and
incorrect substrates, respectively, and that $\omega_{f}^0 >> \Omega_{f}^0$.
In this case, the average velocity of the DNAP, i.e., the average rate 
of polymerization, can be expressed in a MM-like form 
\begin{equation}
V_{p}^{(c)} = \dfrac{\tilde{K}_{cat}[dNTP_{c}]}{\tilde{K}_{M}+[dNTP_c]}\label{m36}
\end{equation}
for the correct nucleotides, where 
\begin{equation}
\tilde{K}_{cat}=\dfrac{\omega_h(\omega_e+\omega_{pr})}{\omega_{pr}+\omega_{e}+\omega_{pf
}}
\end{equation} 
and the effective Michaelis constant is 
\begin{equation}
\tilde{K}_{M}=\dfrac{(\omega_{pr}+\omega_{e})(\omega_{h}+\omega_{r})+\omega_{pf}\omega_{e}}{\omega_{f}^{0}(\omega_{pr}+\omega_{e}+\omega_{pf})}
\end{equation}
Replacing $\omega$ by $\Omega$ and $[dNTP_{c}]$ by $[dNTP_{w}]$ we get the 
average rate of polymerization $V_{p}^{w}$ for the wrong nucleotides. 
In the limit of negligible exonuclease activity, the kinetic diagram 
shown in fig.\ref{fig-3state} reduces to the scheme shown on the left 
panel of fig.\ref{fig-MMlikeP} which is the standard MM-scheme with a single
intermediate complex; in this limit the expressions for $\tilde{K}_{cat}$
and $\tilde{K}_{M}$ are consistent with those for the standard
MM scheme \cite{enzymebook}.

\subsection{Distribution of turnover time for exonuclease}

In this section we derive the DTT for {\it exonuclease} activity of 
the DNAP. We insert a hypothetical state $P_{1}^*$ such that
\begin{equation}
P_{3}\overset{\omega_{e}}{\rightarrow} P_{1}^*\overset{\delta}{\rightarrow} P_{1}
\end{equation}
where in the limit $\delta \rightarrow \infty$, $P_1$ and $P^{*}_{1}$ 
become identical and we recover our original model.\\
For the simplicity of notation, in this subsection we drop the site 
index without loss of any information. 
The master equations for $P_{\mu}(t)$ ($\mu=1,2,3$) and that for 
$P_{1}^*(t)$ are  
\begin{eqnarray}
\dfrac{dP_1 (t)}{dt}&=&-\omega_f P_{1} (t) + (\omega_{r}+\omega_{h}) P_2 (t)
\label{np1}
\end{eqnarray}
\begin{eqnarray}
\dfrac{dP_2 (t)}{dt}&=&\omega_f P_1 (t) - (\omega_r+\omega_{pf}+\omega_h)P_2 (t)\nonumber\\
&+&\omega_{pr} P_3 (t)
\label{np2}
\end{eqnarray}
\begin{equation}
\dfrac{dP_3 (t)}{dt}=\omega_{pf} P_2 (t) - (\omega_e+\omega_{pr})P_3 (t)
\label{np3}
\end{equation}
\begin{equation}
\dfrac{dP_{1}^{*}(t)}{dt}=\omega_{e} P_{3}(t)
\end{equation}
For the calculation of DTT, we impose the initial condition
$P_{1}(0)=1$, and $P_{2}(0)=P_{3}(0)=P_{1}^*(0)=0$. Suppose, $f(t)$ 
denotes the DTT. Then, $f(t) ~\Delta t$ is the probability that one  
exonuxlease cycle is completed between $t$ and $t+\Delta t$, i.e., 
the DNAP was in state 3 at time $t$ and made a transition to the 
state $1*$ between $t$ and $t+\Delta $. Obviously, 
$f(t) \Delta t = \omega_{e} P_{3}(t)$ and, hence,   
\begin{equation}
f(t) = \omega_{e} P_{3}(t) 
\label{m48}
\end{equation}
Using a compact matrix notation, the equations (\ref{np1}),(\ref{np2}) 
and (\ref{np3}) can be written the form
\begin{equation}
 \dfrac{d}{dt}{\textbf{Q}}(t)=\textbf{N}{\textbf{Q}}(t)
 \label{neq}
\end{equation}
where 
\begin{equation}
\textbf{N}=
\begin{bmatrix}
  -\omega_f & \omega_h+\omega_r & 0 \\
   \omega_f\ & -(\omega_h+\omega_r+\omega_{pf}) & \omega_{pr} \\
   0 & \omega_{pf} &-(\omega_e+\omega_{pr}) 
\end{bmatrix}
\end{equation}
and 
\begin{equation}
\textbf{Q}=
\begin{bmatrix}
P_{1}(t) \\  P_{2}(t) \\ P_3(t)
\end{bmatrix}
\end{equation}

\begin{figure}[td]
\begin{center}
\includegraphics[angle=-90,width=0.9\columnwidth]{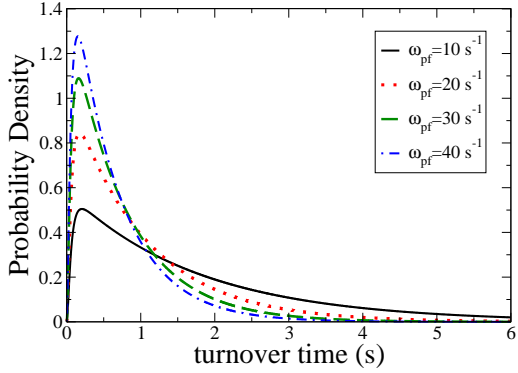}
\end{center}
\caption{(Color online) Probability density of turnover time for 
unproductive exonuclease mode of enzyme, f(t) is plotted for a few 
different values of parameter $\omega_{pf}$. The values of the other 
parameters are (all in $s^{-1}$) $\omega_{f}$=20.0, $\omega_{r}$=5.0,
 $\omega_{pr}$=15.0, $\omega_{h}$=40.0 and $\omega_{e}=4.0$.}
\label{fig5}
\end{figure}

Solution of the equation (\ref{neq}) in Laplace space
\begin{equation}
\tilde{\textbf{Q}}(s)=\textbf{S}(s)^{-1}\tilde{\textbf{Q}}(0) 
\label{m55}
\end{equation}
where
\begin{equation}
 \textbf{S}(s)=s\textbf{I}-\textbf{N}. 
\end{equation} 
Solution (\ref{m55}) for the assumed initial conditions provide,
\begin{equation}
\tilde{P}_{3}(s)=\dfrac{(-1)^{1+3}\mathbf{Co}_{13}}{|\mathbf{S}(s)|}
\end{equation}
which, explicitly in terms of the rate constants, takes the form 
\begin{equation}
\tilde{P}_3(s)=\dfrac{\omega_f \omega_{pf}}{s^3+\alpha 's^2+\beta ' s+\gamma '}
\end{equation}
where
\begin{equation}
\alpha '=\omega_{e}+\omega_{f}+\omega_{h}+\omega_{pf}+\omega_{pr}+\omega_{r}
\end{equation}
\begin{eqnarray}
\beta '&=&\omega_{f}\omega_{pf}+\omega_{e}\omega_{h}
+\omega_{e}\omega_{pf}+\omega_{f}\omega_{pf}+\omega_{f}\omega_{pr}+\omega_{h}\omega_{pr}\nonumber\\ &+&\omega_{e}\omega_{pr}+\omega_{pr\omega_{r}}
\end{eqnarray}
\begin{equation}
\gamma '=\omega_{f}\omega_{pf}\omega_{e}
\end{equation}
Since in the Laplace space the equation (\ref{m48}) becomes 
\begin{equation}
\tilde{f}(s)=\omega_e \tilde{P}_3(s),
\label{lft}
\end{equation}
we get 
\begin{equation}
\tilde{f}(s)=\dfrac{\omega_{e} \omega_f \omega_{pf}}{s^3+\alpha 's^2+\beta ' s+\gamma '}=\dfrac{\omega_{e} \omega_f \omega_{pf}}{(s+\upsilon_{1})(s+\upsilon_{2})(s+\upsilon_{3})}
\label{m60}
\end{equation}
as the DTT in the Laplace space.  
\begin{widetext}

Taking the inverse Laplace transform of equation (\ref{m60}), we get the DTT 
\begin{eqnarray}
f(t) = \biggl[\dfrac{\omega_{f}\omega_{pf}\omega_{e}}{(\upsilon_{1}-\upsilon_{2})(\upsilon_{1}-\upsilon_{3})}\biggr]e^{-\upsilon_{1}t}
+\biggl[\dfrac{\omega_{f}\omega_{pf}\omega_{e}}{(\upsilon_{2}-\upsilon_{1})(\upsilon_{2}-\upsilon_{3})}\biggr]e^{-\upsilon_{2}t}
+\biggl[\dfrac{\omega_{f}\omega_{pf}\omega_{e}}{(\upsilon_{3}-\upsilon_{1})(\upsilon_{3}-\upsilon_{2})}\biggr]e^{-\upsilon_{3}t}
\end{eqnarray}
where $\upsilon_{1}$,$\upsilon_{2}$,$\upsilon_{3}$ are solution of the following equation
\begin{eqnarray}
& &\upsilon^{3}-(\omega_{e}+\omega_{f}+\omega_{h}+\omega_{pf}+\omega_{pr}+\omega_{r})\upsilon^{2}+(\omega_{f}\omega_{pf}+\omega_{e}\omega_{h}
+\omega_{e}\omega_{pf}+\omega_{f}\omega_{pf}+\omega_{f}\omega_{pr}+\omega_{h}\omega_{pr}+\omega_{e}\omega_{pr}+\omega_{pr\omega_{r}})\upsilon
\nonumber \\
&-&(\omega_{f}\omega_{pf}\omega_{e}) = 0.
\end{eqnarray}
\end{widetext}
This DTT is plotted in fig(\ref{fig5}). Plots are consistent with the 
intuitive expectation that increasing $\omega_{pf}$ leads to a decrease 
the turnover time.

Suppose $\langle t \rangle$ denotes the mean time gap between the
completition of the successive {\it exonuclease} reactions catalyzed
by the DNAP. Then the average rate $V_{e} = 1/\langle t \rangle$ of
the exonuclease reaction can be expressed in a MM-like form \cite{kou05}
\begin{equation}
V_{e}^{(c)}=\dfrac{K_{cat}[dNTP_c]}{K_{M}+[dNTP_c]} ~{\rm with}~
K_{cat}=\dfrac{\omega_{pf}\omega_{e}}{\omega_e+\omega_{pf}+\omega_{pr}}
\label{eq-vec}
\end{equation}
for the correct nucleotides where $K_{M} = \tilde{K}_{M}$.
Replacing $\omega$ by $\Omega$ and $[dNTP_c]$ by $[dNTP_w]$ in (\ref{eq-vec})
we get $V_{e}^{(w)}$ for the wrong nucleotides.
In the limit $\omega_h \to 0$, $\omega_{pr} \to 0$, the kinetic diagram
shown in fig.\ref{fig-3state} reduces to the simpler scheme shown on the
right panel of fig.\ref{fig-MMlikeP} which is essentially a
generalized MM-like scheme with two intermediate states. Not
surprisingly, in this limit, the average rate of the exonuclease
reaction is consistent with that of the MM-like scheme with two
intermediate states \cite{enzymebook}.

\subsection{Quantitative measures of error}

\begin{figure}[t]
\begin{center}
\includegraphics[angle=-90,width=0.9\columnwidth]{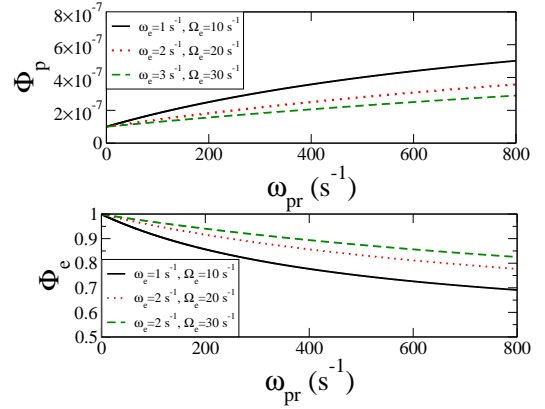}
\end{center}
\caption{(Color online) $\Phi_{p}$ and $\Phi_{e}$ plotted against 
$\omega_{pr}$ while the ratio $\Omega_{pr}/\omega_{pr}=0.1$ is kept fixed. 
The three curves correspond to $\omega_{e} = 1.0$s$^{-1}$, $2.0$s$^{-1}$, 
$3.0$s$^{-1}$.  The values of the other pararemeters are (all in s$^{-1}$):
$\omega_{f} = 1.0$,       $\Omega_{f} = 10^{-5}$,
$\omega_{pf} = 0.1$,           $\Omega_{pf} = 10.0$,
$\omega_{r} = .1$,          $\Omega_{r} = 1.0$,
$\omega_{h} = 10.0$,             $\Omega_{h} = 0.1$.
}
\label{fig-errors}
\end{figure}

Note that
$\Phi_{p} = \tilde{V}_{p}^{(w)}/(\tilde{V}_{p}^{(w)}+\tilde{V}_{p}^{(c)})$
is the fraction of nucleotides {\it misincorporated} in the final
product of replication.  Similarly, the fraction
$\Phi_{e} = V_{e}^{c}/(V_{e}^{c}+V_{e}^{w})$
is a measure of the {\it errorneous severings}, i.e., fraction of the
cleaved nucleotides that were incorporated correctly into the growing DNA.
Since $\omega_{pr}$ is the strength of the ``coupling'' between the two
different enzymatic activities, we plot $\Phi_{p}$ and $\Phi_{e}$ against
$\omega_{pr}$ in fig.\ref{fig-errors} for a few typical sets of values of
the model parameters.

Decreasing $\Phi_{e}$ with increasing $\omega_{pr}$ is a consequence
of the escape route via $\omega_{pr}$ for the correctly incorporated
nucleotides that get transferred unnecessarily to the exonuclease site.
It is the increasing number of such correctly incorporated nucleotides
recused from the exonuclease site that leads to the lowering of
$\Phi_{p}$ with increasing $\omega_{pr}$. The limiting values of
$\Phi_{p}$ and $\Phi_{e}$ in the limit of large $\omega_{pr}$, are
determined by the corresponding limiting expressions
${\tilde V}_{p}^{(c)} \simeq (\omega_{f}\omega_{h})/(\omega_{h}+\omega_{f}+\omega_{r})$,
and
$V_{e}^{(c)} \simeq (\omega_{f}\omega_{pf}\omega_{e})/[\omega_{pr}(\omega_{h}+\omega_{f}+\omega_{r})]$.
(Expressions for ${\tilde V}_{p}^{(w)}$ and $V_{e}^{(w)}$ are similar
in the limit of large $\Omega_{pr}$.)

\section{Summary and conclusion}

Here we have theoretically investigated the effects of the coupling
of two different modes of enzymatic activities of a DNAP, in one of
these it elongates a DNA whereas in the other it shortens the same
DNA. The fundamental questions we have addressed here in the context
of DNA replication have not been addressed by earlier theoretical
works \cite{goel03}.
The effects of tension on the polymerase and exonuclease activities,
which have been the main focus of the earlier works \cite{goel03},
will be reported elsewhere \cite{sharma12}.
The mechanism of error correction by DNAP is somewhat different from
the mechanism of transcriptional proofreading which is intimately
coupled to ``back tracking'' of the RNA polymerase \cite{voliotis09,sahoo11}.

We have derived exact analytical formulae for the cDTD and DTT which
will be very useful in analyzing experimental data in single DNAP
biophysics, particularly its stepping patterns and enzymatic turnover.
In spite of their coupling, the average rates of both the enzymatic
activities are MM-like; the analytical expressions for the effective
MM parameters explicitly display the nature of the coupling of the
two kinetic processes. We have also reported exact analytical
expressions for the fractions $\Phi_{p}$ and $\Phi_{e}$ which measure
{\it replication error} and {\it erroneous cleavage}; these expressions
can be used for analyzing data from both single molecule \cite{ibarra09}
and bulk \cite{wong91} experiments on wild type and mutant DNAPs. \\

{\bf Acknowledgements}: We thank Stefan Klumpp for useful comments.
This work has been supported at IIT Kanpur by the Dr. Jag Mohan Garg 
Chair professorship (DC) and a CSIR fellowship (AKS). This research (DC) 
has been supported in part also by the Mathematical Biosciences Institute 
at the Ohio State University and the National Science Foundation under 
grant DMS 0931642.

\section{Bibliography}

\end{document}